# Monte Carlo calculation of the organ equivalent dose and effective dose due to immersion in a $^{16}$N beta source in air using the ICRP Reference Phantoms

José M. Gómez-Ros[1], Montserrat Moraleda[1], Pedro Arce[1], Duc-Ky Bui[2], Thi-My-Linh Dang[2], Laurent Desorgher[3], Han Sung Kim[4], Dragana Krstic[5], Michał Kuć[6], Ngoc-Thiem Le[2], Yi-Kang Lee[7], Ngoc-Quynh Nguyen[2], Dragoslav Nikezic[5,8], Katarzyna Tymińska[6], Tomas Vrba[9]

[1] CIEMAT, Av. Complutense 40, 28040 Madrid, Spain

[2] Institute for Nuclear Science and Technology, 179 Hoang Quoc Viet St., Hanoi, Vietnam

[3] CHUV, Institute of Radiophysics, Rue du Grand-Pré 1, 1007 Lausanne, Switzerland

[4] Korea Institute of Radiological and Medical Sciences (KIRAMS), Seoul, South Korea

[5] Univ. of Kragujevac, Faculty of Science, Radoja Domanovica 12, 34000 Kragujevac, Serbia

[6] National Centre for Nuclear Research, ul. A. Soltana 7, 05-400 Otwock, Poland

[7] Université Paris-Saclay, CEA, Service d'Études des Réacteurs et de Mathématiques Appliquées, 91191, Gif-sur-Yvette, France

[8] State University of Novi Pazar, Vuka Karadžića 66, 36300 Novi Pazar, Serbia

[9] Faculty of Nuclear Sciences and Physical Engineering, Czech Technical University in Prague, Prague, Czech Republic

**Abstract:** This work summarises the results of a comparison organized by EURADOS focused on the usage of the ICRP Reference Computational Phantoms. This activity aimed to provide training for the implementation of voxel phantoms in Monte Carlo radiation transport codes and the calculation of the dose equivalent in organs and the effective dose. This particular case describes a scenario of immersion in a $^{16}$N beta source distributed in the air of a room with concrete walls where the phantom is located. Seven participants took part in the comparison of results using GEANT4, TRIPOLI-4 and MCNP family codes, and there was detected a general problem when calculating the dose to skeletal tissue and the remainder tissue. After a process of feedback with the participants the errors were corrected and the final results reached an agreement of ±5%.

# 1. Introduction

In recent years, comparison exercises and benchmark studies have been organized within the activities of the European Radiation Dosimetry Group EURADOS (Gómez-Ros et al., 2008; Broggio et al., 2012; Vrba et al., 2014; Nogueira et al., 2015; Vrba et al., 2015) on the use of Monte Carlo simulation codes and particular advanced tools to share knowledge, detect difficulties and improve the methods and approaches employed in the field of dosimetry.

The International Commission on Radiological Protection (ICRP) adopted the voxel (volume elements) models as the base for its computational phantoms to represent the Reference Adult Male and Female (ICRP, 2009). These phantoms are very realistic, since they have been derived from medical image data of real persons, and they can be used together with different radiation transport codes to simulate the transport of radiation through the tissues and to evaluate the energy deposition in body organs and tissues due to external or internal sources of radiation.

To investigate the usage of these phantoms for dosimetry applications, EURADOS has organized this training activity in which voluntary participants are invited to solve specific tasks of practical interest in occupational, environmental and medical dosimetry (Zankl et al., this issue). The aim was to look into the procedure of phantoms implementation in different radiation transport codes, to give participants the opportunity to check their own calculations against quality-assured solutions, and to improve their approach, if needed. Several exposure scenarios have been proposed to deal with different radiation sources (photons, electrons and neutrons) and different irradiation geometries: point sources, ground and air contamination, X-ray examinations and internal dose assessment.

This work addresses a scenario with an anthropomorphic phantom immersed in a $^{16}$N beta source homogenously distributed in the air inside a closed room with concrete walls. The $^{16}$N is a neutron activation product of coolant water in fission and fusion reactors. Its beta spectrum has endpoint of 10.4 MeV so producing high enough energetic particles to test the electron-photon-positron transport calculation of modern Monte Carlo codes. The phantom is located at the centre of the room and the aim of the exercise is to calculate equivalent doses for both adult phantoms, male and female, and the resulting effective dose. A dedicated template for the answers was distributed to the participants to provide information about the calculation of the quantities and to assure the homogeneity of the results and their uncertainties. The solutions submitted by the participants were compared with the reference solution and different problems were detected and analysed.

## 2. Materials and methods

*2.1. ICRP Reference Phantoms*

The ICRP developed two adult phantoms, male and female (ICRP, 2009) to be used as reference computational phantoms for radiological protection purposes. They are based on real individuals but modifying their models to fulfil the anatomical characteristics, dimensions and masses, of the Reference Male and Female subjects (ICRP, 2002).

A total of 141 organs and tissues have been identified in the computational phantoms (including air inside the phantom). The list of the organs with their identification number (ID), medium, density and mass are reported in Annex A of the ICRP publication 110 (ICRP, 2009) and the elemental compositions of the 53 different tissues of the phantoms (air included) are given in Annex B in the same publication and are accessible in separate data files enclosed with the publication.

The Reference Adult Male Phantom represents an individual with height 176 cm and mass 73 kg and consists of a three-dimensional voxel array arranged in 254 columns (x co-ordinate), 127 rows (y co-ordinate) and 222 slices (z co-ordinate). The file has an ASCII format and it is named AM.dat. The external parallelepiped dimensions of the voxel phantom are $54.2798 \times 27.1399 \times 177.6$ cm$^3$ with a voxel size of $0.2137 \times 0.2137 \times 0.8$ cm$^3$.

The file AF.dat contains the data of the Reference Female Phantom and corresponds to a woman with height 163 cm and mass 60 kg. The array dimensions are $299 \times 137 \times 348$ columns, rows and slices, respectively with external phantom dimensions of $53.0725 \times 24.3175 \times 168.432$ cm$^3$. The voxel size of the female phantom is $0.1775 \times 0.1775 \times 0.484$ cm$^3$.

*2.2. Irradiation geometry*

The Reference Phantom is within a $600 \times 600 \times 400$ cm$^3$ room made of concrete walls, floor and ceiling 50 cm thick. The room is filled with air and the phantom is on the floor at the centre of the room (Figure 1). The $^{16}$N source is uniformly distributed in air and its beta energy spectrum was given as a density distribution according to ICRP Publication 107 (ICRP, 2008). The beta decay of the $^{16}$N is accompanied of a gamma emission but this exercise considers exclusively the exposure to the beta source.

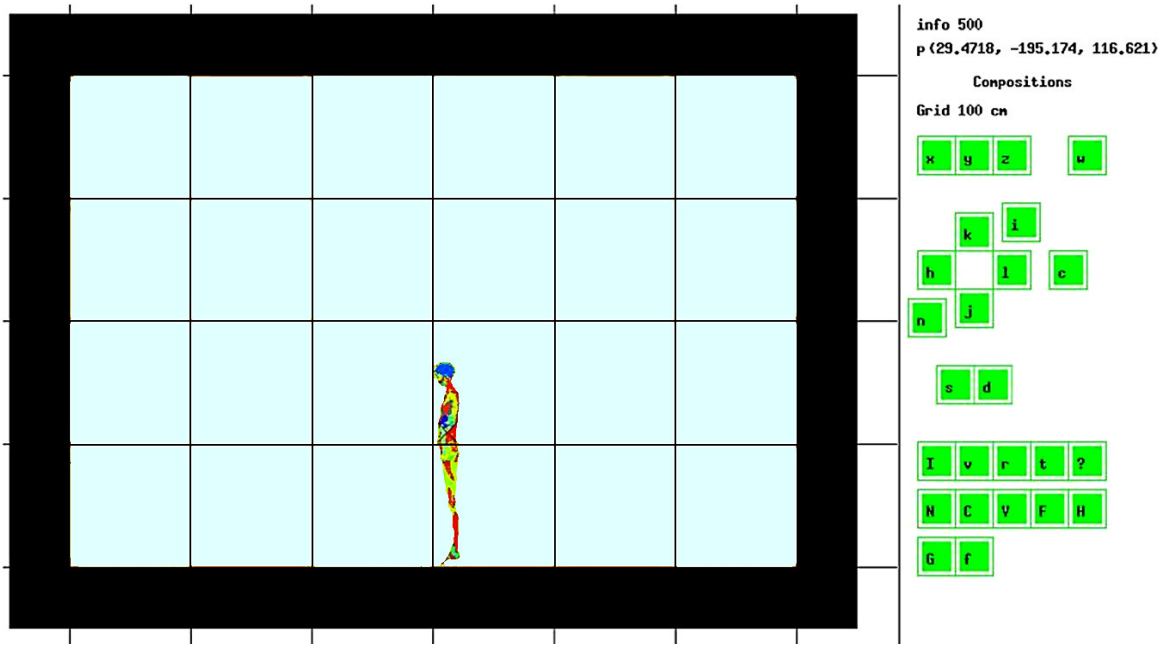

**Figure 1.** Visualization with TRIPOLI-4 of the proposed geometry. The phantom stands in the centre of a room of size 600x600x400 cm$^3$ with concrete walls 50 cm thick.

*2.3. Requested results*

Two main tasks were proposed in this work to study the use of the phantoms for the calculation of some dosimetry quantities. A first task consisted of calculating the equivalent dose rates per activity concentration in air, in units of (Sv s$^{-1}$)/(Bq m$^{-3}$) in all the phantom indexed parts. The equivalent dose ($H_T$) in an organ or tissue T is calculated from the mean absorbed dose (D$_{T,R}$) as: $H_T = \sum_R w_R D_{T,R}$, where $w_R$ is the radiation weighting factor for radiation R. Because the radiation weighting factor for electrons and photons is $w_R$=1, the equivalent dose is directly calculated as the energy deposited in the organ divided by the organ mass.

The second task asked to calculate the equivalent doses in specific organs: bladder, bone surface, breast, colon, gonads, liver, lung, oesophagus, red bone marrow, skin, stomach, thyroid and remainder tissues, required to obtain the effective dose. Most of these organs are composed of several parts identified with different ID number in the segmentation process of the phantoms creation and the equivalent dose is the total energy deposited divided by the total organ or tissue mass.

In general, separate blood content should not be added to the organ and tissue masses if they are not identified as a separate organ, such as lungs which comprise voxel organs 96, 97, 98 and 99 (lung tissue and blood) (ICRP, 2009).

For the dosimetry calculations in red bone marrow and endosteum (bone surface), it was proposed to follow the method described in ICRP 116 (ICRP, 2010). The absorbed dose to active marrow, $D_{skel}(AM)$, and the absorbed dose to endosteum, $D_{skel}(TM50)$, are calculated as follows:

$$D_{skel}(AM) = \frac{\sum m(AM,x)}{m(AM)} D(SP,x) \qquad (1)$$

$$D_{skel}(TM50) = \frac{\sum m(TM50,x)}{m(TM50)} D(SP,x) + \frac{\sum m(TM50,x)}{m(TM50)} D(MM,x) \qquad (2)$$

where $m(AM,x)$ and $m(TM50,x)$ are respectively the mass of active bone marrow and mass of endosteum in bone site x, according to Table 3.2 in (ICRP, 2010); $m(AM)$ and $m(TM50)$ are the masses of active marrow and endosteum summed across the entire skeleton; $D(SP,x)$ and $D(MM,x)$ are the absorbed dose to spongiosa and to medullary cavities in bone site x, respectively.

According to ICRP recommendations (ICRP, 2007), the absorbed dose in remainder tissues is calculated as the arithmetic mean dose of 13 organs: adrenals, extrathoracic region, gall bladder, heart, kidneys, lymphatic nodes, muscle, oral mucosa, pancreas, small intestine, spleen, thymus and prostate/uterus.

Finally, it was asked to evaluate the effective dose rate per activity concentration of $^{16}$N in air (Sv s$^{-1}$)/(Bq m$^{-3}$) that is calculated as a sex-averaged of the equivalent doses assessed for organ or tissue T of the Reference Male and Reference Female according to:

$$E = \sum_T w_T \left[ \frac{H_T^{male} + H_T^{female}}{2} \right] \qquad (3)$$

where $H_T^{male}$ and $H_T^{female}$ are equivalent doses for tissue T in male and female phantom and $w_T$ is the tissue weighting factor for organ T recommended in (ICRP, 2007).

*2.4. Monte Carlo codes*

Geant4 (Agostinelli et al., 2003; Allison et al., 2016) is a free software package toolkit that makes use of object-oriented programming. In this work, the version 10.4 was used, with the so-called Shielding physics list that considers the Geant4 standard electromagnetic physics models down to 1 keV low energy limit. Cut in range of 0.5 mm was selected in the simulation, thus only secondary particles with range above the threshold were transported.

GAMOS (Arce et al., 2014) is a Geant4 based framework that permits to implement the most common requirements of a Medical Physics application without any need of C++ programming. The selected physics is the so-called "GmEMPhysics" with the electromagnetic "low-energy" physics models based on the sampling of the Livermore databases EEDL for electrons (Seltzer et al., 1989) and EADL for photons (Perkins et al., 1991). For the simulations, a cut in range of 0.01 mm was selected.

MCNP5 (X-5 Monte Carlo Team, 2003a; X-5 Monte Carlo Team, 2003b), MCNPX 2.7 (Pelowitz, 2011) and MCNP6 (Werner, 2017) are different versions of the MCNP family codes. Two photon cross section libraries have been used: mcplib04 (White, 2003) and mcplib84 (White 2012), that are identical but for the correction in the most recent version of a bug related to data format when the Doppler broadening is used. The electron cross-section data library el03 (Adams, 2000; X-5 Monte Carlo Team, 2003a) has been used in all the cases.

TRIPOLI-4® (version 11, November 2018) is the fourth generation of the Monte Carlo radiation transport code developed at CEA/Saclay (Brun et al., 2015). The official data library for applications, named CEAV5.1.1, is mainly based on the JEFF-3.1.1 for neutron and the ENDL-97 for photon and electron (Mancusi et al., 2018). The code sets the cut-off energy at 1 keV and the default value of thin_layer_substep was used to allow a more rigorous processing of frontier approaches.

*2.5. The reference solution*

The reference solution was calculated with MCNPX2.7 using the libraries mcplib84 and el03 and default cut-off energies (1 keV). In all the cases, $10^{10}$ source particles were simulated in order to get results with relatively low statistical uncertainties. The statistical error expressed as relative standard deviation was below ±3% for the majority of the targets.

To convert voxel phantoms into MCNPX syntax, the data files AM.dat and AF.dat were processed with an ImageJ (Schneider et al., 2012) plug-in specially developed (Gomez-Ros et al., 2007). The feature of repeated structures has been used to describe the geometry as a lattice of cubic cells that fill the space. Each cell (universe) is associated to an organ through its ID number and it is filled with the appropriate material.

*2.6. Voxel phantom implementation in the different MC codes*

The participants used different procedures to handle the voxel phantoms depending on the MC code. For Geant4, the phantoms were implemented in two ways. In one case, the phantom parameterization and constructor were adapted by modifying the Geant4 DICOM medical examples in order to read the ICRP voxel phantom data and set the corresponding voxel dimensions and materials. Specified tallies were implemented into Geant4 allowing to register the absorbed dose and track length of particles in group of voxels representing the organs. For the bone dosimetry, the photon fluence crossing the red bone marrow and the endosteum were computed by dividing the corresponding track length by the organ volume, while the absorbed doses were computed from the photon fluence by using fluence to dose conversion factors published by the ICRP.

Alternatively, the ICRP110 data were also converted to Geant4 text file voxel phantom format used in the official Geant4 example extended/medical/DICOM with an ad-hoc C++ code. Each voxel has been assigned to an organ using the information contained in AF/AM.dat. The density has been assigned as given in the file AF/AM_organs.dat and the material composition as given in the file AM/FM_media.dat. The tracking of particles has been done with the G4RegularNavigation algorithm (Arce et al., 2008), skipping the frontiers of contiguous voxels which have the same material.

The voxel phantoms were implemented into the MCNP codes (MCNPX, MCNP5, MCNP6) using similar procedures based in the repeated structures feature to describe a repetitive geometry. Each organ ID was represented by a volume element with the size of the voxel phantom resolution and it is repeated using the LAT card to create a 3-D lattice. A unique universe number was assigned to each organ, and the universe numbers were filled with the each 3-D lattice cell to construct the phantom geometry.

Processing the voxel phantom data to create the MCNP input file were done using specific-purpose programmes: i) R scripts (R Core Team, 2013) to read and transform the phantom data to MCNP5 format, producing four files: universe, lattice, material and tally, and the rest of the input was manually implemented to include the surfaces and the source; ii) a Python code (Van Rossum and Drake Jr, 1995), to convert the phantom data into MCNPX format; iii) a home-made script written in Fortran90 to process data, first counting the number of repetitions and then rewriting with some additional editing; and iv) another home-made code to incorporate the phantom into MCNP6.

For TRIPOLI-4, the lattice geometry was validated in previous studies for criticality-safety analyses and core physics calculations (Lee, 2003; Lee and Hugot, 2009; Lee, 2015) using the lattice operators 'EXCEPT' and 'KEEP'. In a similar way, the modelling method was used to describe the void voxels and the organ-tissue voxels for the phantoms. To verify the TRIPOLI-4 modelling, the T4G graphical tool was helpful to check the organ-tissue positions, organs' model and dimensions, source-phantom locations, and associated media names. In order to accelerate the navigation and display from the phantom's environment to organ views in a fast and interactive way the T4G tool was used in parallel mode (Hugot and Lee, 2011; Lee, 2018). In this study, the electron-photon-electron-positron cascade showers option of TRIPOLI-4 was turned on.

## 3. Results and discussion

Seven participants from different countries took part in the exercise using different Monte Carlo codes to perform the calculations (GEANT4, GAMOS, MCNP/MCNPX, TRIPOLI-4, as it is summarized in Table 1.

**Table 1.** Summary of used MC codes and cross-section libraries.

| MC code | photon cross-section library | electron cross-section library |
|---|---|---|
| GEANT4 10.04 | EPDL | EEDL |
| GEANT4 10.04 / GAMOS 6.0 | EPDL | EEDL |
| MCNPX 2.7 | mcplib84 | el03 |
| MCNP5 v1.60 | mcplib84 | el03 |
| MCNP6.2 | mcplib04 | el03 |
| MCNP6.2 | mcplib84 | el03 |
| TRIPOLI-4 | EPDL97 | EEDL + Bremsstrahlung |

*3.1. Dose in all the phantom indexed parts*

Participants provided the results for the equivalent doses and effective dose together with the associated uncertainties, and their solutions were compared with the reference one. When a difference with the reference solution higher than ±10% was obtained, the results were examined closely and participant was asked to revise the calculation looking for the causes of the discrepancy.

The first set of data consisted of the equivalent dose rates for each of the 139 indexed parts of the phantoms (from ID=1 to 139) excluding the air inside the phantom and the skin at top and bottom of the phantom. From the point of view of the dosimetry methodology, those calculations are simple and they allow to detect mistakes related with the geometry, the source or the simulation of the radiation transport.

In order to rule out an incorrect construction of the phantom a check of the organ volumes based on a stochastic estimation of the fluence in the regions of interest was carried out. Most of the participants managed to implement the voxel phantoms correctly in their MC codes and only one of them detected an error in handling the phantom geometry. Nevertheless, from the 7 answers only 2 participants initially provided correct results for the $H_T$ in the 139 organs defined in the phantom.

Some of the mistakes were due to incorrect source definition, the energy spectrum and/or the spatial distribution. Particularly common among the participants has been the missing of the source in the air surrounding the body, but still inside the limits of the voxel phantom, resulting in an incorrect spatial distribution of the source. This fact is important since it can contribute to the absorbed dose around ±10% of the total dose in some organs. In addition, the beta energy spectrum of $^{16}N$ given as a probability density function was converted to probability histogram by some participants resulting in a slightly different emission especially at higher energies which results in an increase of 10-15% in the absorbed dose in some organs.

Other discrepancies were related with the Monte Carlo method, in particular, low statistic uncertainties. Electrons up to several MeV have a relative short mean free path in human tissues and are quickly attenuated so a large number of particles needed to be simulated to get reliable results which is a handicap in terms of computer time. Also, systematic mistakes related to the normalisation of the results according to the volume of the source were found.

*3.2. Equivalent dose in selected organs and effective dose*

The second set of data included the equivalent dose rates in 13 organs (bladder, bone surface, breast, colon, gonads, liver, lung, oesophagus, red bone marrow, skin, stomach, thyroid and remainder tissues) and the effective dose. From these results, the approaches and methods to get those quantities could be analysed since the calculations need some treatment of the Monte Carlo results. The results of equivalent dose rates per activity concentration are depicted in Figure 2 and they show big discrepancies.

The sources of errors due to the calculation method were: incorrect use of the skeletal fluence-to-dose response functions for photons when the problem dealt with electrons; errors in the mass values of skeletal tissues (red bone marrow and bone surface) used in equations 1 and 2; wrong calculation of absorbed dose in organs consisting of several parts with different organ ID in the phantom description; and wrong calculation of the absorbed dose for remainder tissues. Also improper values of $w_T$ factors were detected in the calculation of the effective dose.

After a discussion and revision process with the participants, the amended results for the organ equivalent dose rates were in good agreement with the reference solution, with differences less than ±5% in most cases, as it can be seen in Figure 3. Two results still deviate from the reference values (nearly ±10% for some organs) due to lack of time to repeat all the simulations after the deadline for completion of the comparison (some simulations require a long computing time and multiprocessing capabilities were not available for everybody). Those results still have some problems with the source definition and the skeletal dose calculation.

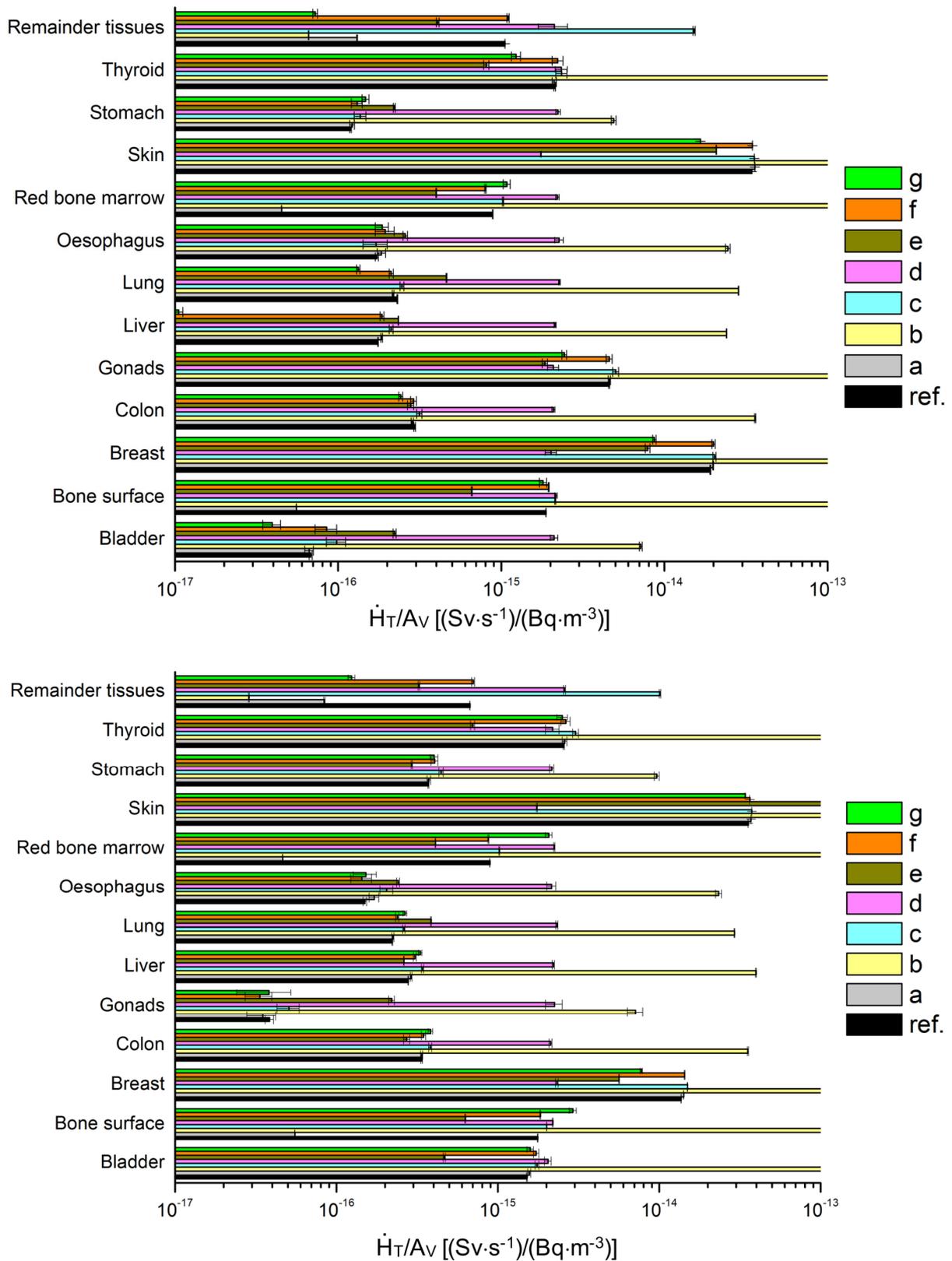

**Figure 2.** Equivalent dose rate per $^{16}$N activity concentration, $\dot{H}_T/A_V$, for selected organs before correction of the results: a) the male phantom; b) female phantom.

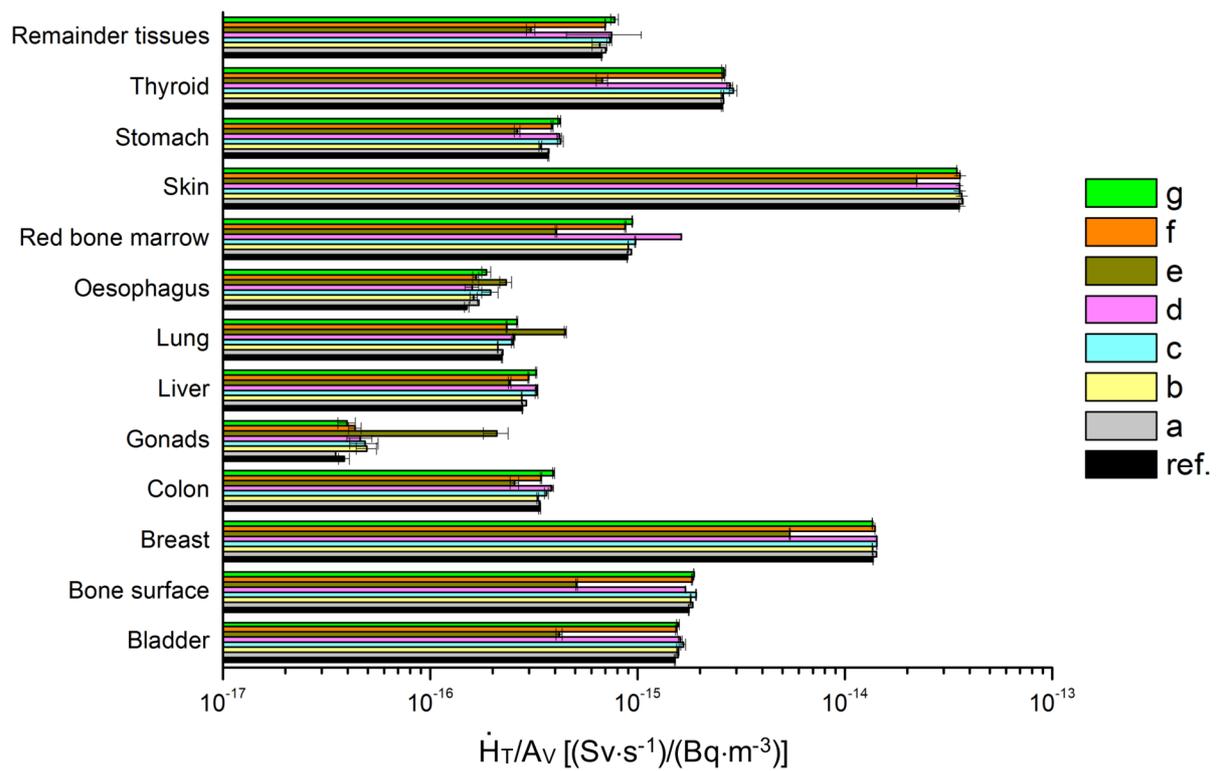

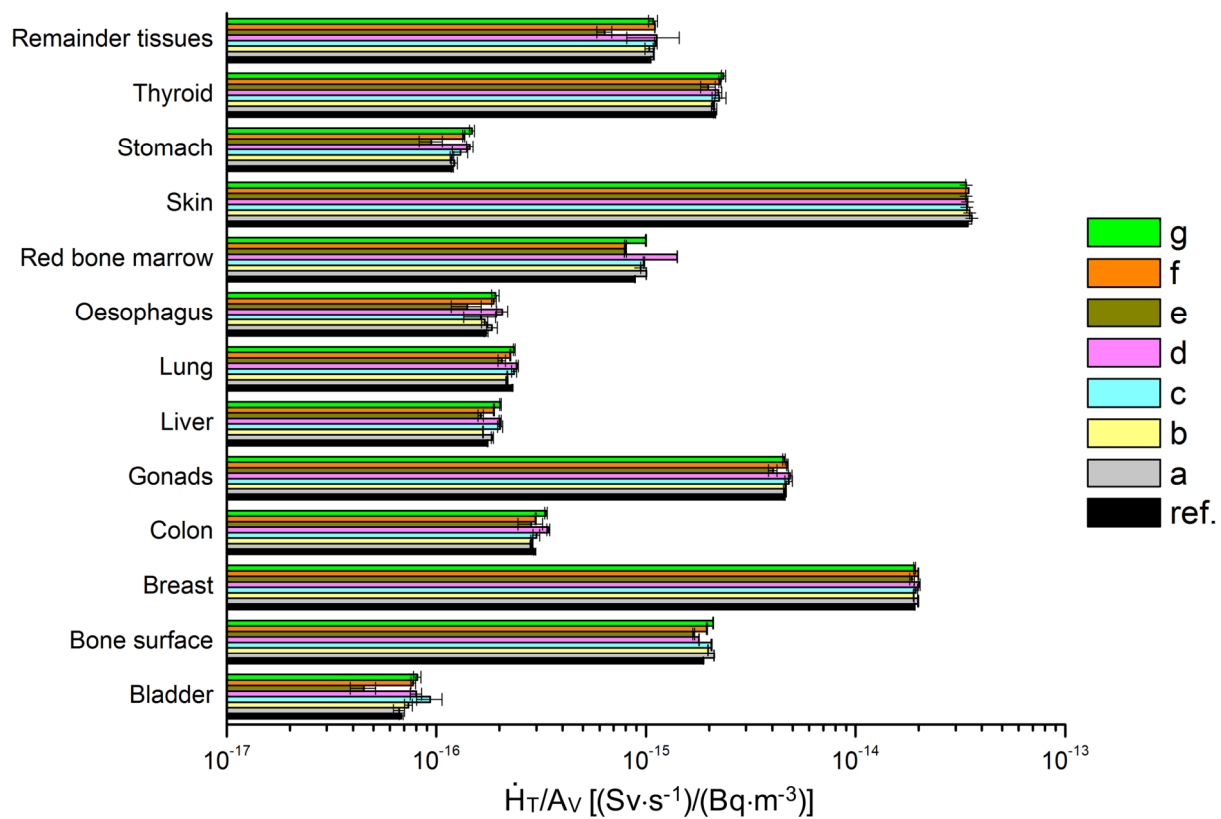

**Figure 3.** Equivalent dose rate per $^{16}$N activity concentration, $\dot{H}_T/A_V$, for selected organs after correction of the detected mistakes: a) the male phantom; b) female phantom.

A general problem in the estimation of the equivalent dose in skeletal tissue and remainder tissues according to ICRP recommendations have been noticed. The contribution of each organ to the effective dose is shown in Figure 4. Although the main contributions are the breast, skin and gonads, the skeletal tissue represents around 4% and the remainder tissue another 3% so they are not negligible terms.

The calculated values of effective dose rate per $^{16}$N activity concentration, $\dot{E}/A_V$, once the detected mistakes have been corrected are shown in Table 2. Since these values have been calculated considering only the beta emission of $^{16}$N for immersion in a cloud inside the room depicted, it is not comparable with the coefficient provided in ICRP Publication 144 (ICRP, 2020a).

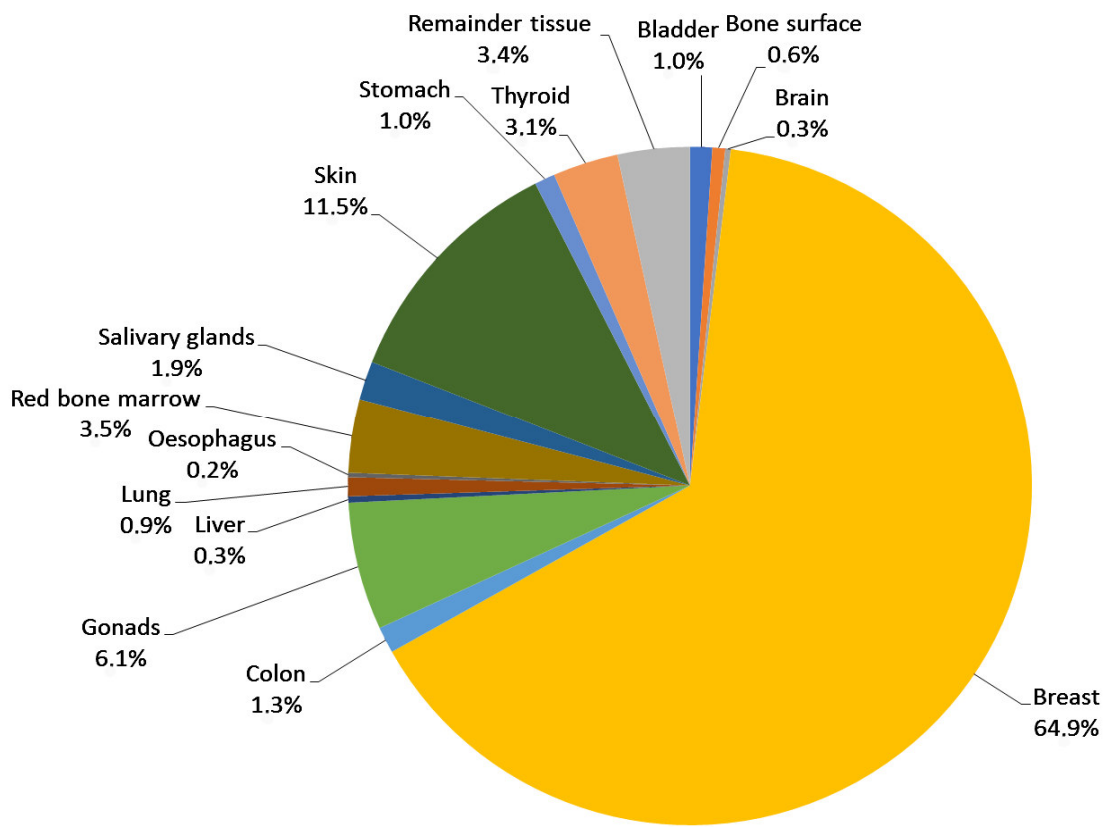

**Figure 4.** Contribution of each organ to the calculated effective dose.

**Table 2.** Comparison of the calculated values of effective dose rate per $^{16}$N activity concentration, $\dot{E}/A_V$, after correction of the detected mistakes.

| participant | $\dot{E}/A_V$ (Sv·s$^{-1}$)/(Bq·m$^{-3}$) |
|:---:|:---:|
| reference | 3.03×10$^{-15}$ |
| a | 3.14×10$^{-15}$ |
| b | 2.96×10$^{-15}$ |
| c | 3.13×10$^{-15}$ |
| d | 3.19×10$^{-15}$ |
| e | 2.80×10$^{-15}$ |
| f | 3.11×10$^{-15}$ |
| g | 3.16×10$^{-15}$ |

## 4. Conclusions

The goal of this exercise was to study the implementation and usage of the ICRP reference computational phantoms together with modern Monte Carlo transport codes and it was part of an EURADOS comparison dealing with different situations of the radiation protection field.

This exercise provides training for the correct use of computational reference phantoms and the calculation of quantities of interest. The comparison gives the participants the opportunity to compare the calculation methods and to improve their approaches. The success of the work relies in a continuous and fluid communication among participants and organizers to detect specific problems.

Concerning the voxel phantom construction and despite some errors in tissue compositions, only one participant had problems with the phantom geometry. In general, the participants managed to implement the phantoms in their codes without difficulties and only a few problems related with the normalization of results and the source definition were found.

Most of the problems arose from the method to estimate absorbed doses to the bone and the remainder tissue described in ICRP116. The method to calculate $H_T$ needed some advice and further calculations in all cases.

After analysis of results and subsequent corrections, participants calculated $H_T$ within a ±5% agreement with the reference value in most cases, even though the discrepancies were up to ±10% in small organs.

Future activities are foreseen on the use of the mesh-type reference computational phantoms released by ICRP (ICRP, 2020b) that overcome some limitations of the voxel geometry mainly due to resolution of small tissue structures.

**Acknowledgements**

This work has been supported by EURADOS (European Radiation Dosimetry Group), within the activities of Working Group 6: Computational Dosimetry.